\documentclass{PoS}

\title{The Air Microwave Yield (AMY) experiment - A laboratory measurement of the microwave emission from extensive air showers}

\ShortTitle{The Air Microwave Yield (AMY) experiment}

\author{J. Alvarez-Mu\~niz$^{1}$, M. Blanco$^{2}$, M. Boh\'a\v cov\'a$^{3}$,
B. Buonomo$^{4}$, G. Cataldi$^{5}$, M. R. Coluccia$^{5,6}$, P. Creti$^{5}$, I.
De Mitri$^{5,6}$, C. Di Giulio$^{7}$, P. Facal San Luis$^{8}$, L.
Foggetta$^{4}$, R. Ga\"ior$^{2}$, D. Garcia-Fernandez$^{1}$,
M. Iarlori$^{9}$, S. Le Coz$^{10}$, A. Letessier-Selvon$^{2}$, \speaker{K. Louedec}$^{10}$,
I. C. Mari\c{s}$^{2}$, D. Martello$^{5,6}$, G. Mazzitelli$^{4}$, M. Monasor$^{8}$, L. Perrone$^{5,6}$, R. Pesce$^{11}$,
S. Petrera$^{9}$, P. Privitera$^{8}$, V. Rizi$^{9}$, G. Rodriguez Fernandez$^{7}$, F. Salamida$^{12}$, 
G. Salina$^{7}$, M. Settimo$^{2}$, P. Valente$^{4}$, J. R. Vazquez$^{14}$, V. Verzi$^{7}$, C. Williams$^{8}$\\

$^1$ Depto. de Fisica de Particulas, Universidad de Santiago de Compostela, Spain \\
$^2$ Laboratoire de Physique Nucl\'eaire et de Hautes Energies (LPNHE),
Universit\'es Paris 6 et Paris 7, CNRS-IN2P3, Paris, France \\
$^3$ Institute of Physics, Academy of Sciences of the Czech Republic, Prague, Czech Republic \\
$^4$ Istituto Nazionale di Fisica Nucleare - Laboratori Nazionali di Frascati, Frascati, Italy \\
$^5$ Sezione INFN, Lecce, Italy \\
$^6$ Dipartimento di Matematica e Fisica Ennio De Giorgi, Universit\`{a} del Salento, Lecce, Italy \\
$^7$ Sezione INFN, Roma Tor Vergata, Italy \\
$^8$ University of Chicago, Enrico Fermi Institute Kavli Institute for
Cosmological Physics, Chicago, USA \\
$^9$ Dipartimento di Fisica, Universit\`{a} dell'Aquila and sezione INFN,
l'Aquila, Italy   \\
$^{10}$ Laboratoire de Physique Subatomique et de Cosmologie (LPSC), UJF-INPG, CNRS-IN2P3, Grenoble, France \\
$^{11}$ Dipartimento di Fisica dell'Università and INFN, Genova, Italy \\
$^{12}$ Institut de Physique Nucl\'{e}aire d'Orsay (IPNO), Universit\'{e} Paris
11, CNRS-IN2P3, France  \\
$^{14}$ Universidad Complutense de Madrid, Madrid, Spain \\

        E-mail: \email{karim.louedec@lpsc.in2p3.fr}}

\abstract{The AMY experiment aims to measure the microwave bremsstrahlung radiation (MBR) emitted by air-showers secondary electrons accelerating in collisions with neutral molecules of the atmosphere. The measurements are performed using a beam of 510 MeV electrons at the Beam Test Facility (BTF) of Frascati INFN National Laboratories. The goal of the AMY experiment is to measure in laboratory conditions the yield and the spectrum of the GHz emission in the frequency range between 1 and 20 GHz. The final purpose is to characterise the process to be used in a next generation detectors of ultra-high energy cosmic rays. A description of the experimental setup and the first results are presented.}

\FullConference{The European Physical Society Conference on High Energy Physics \\
		 18-24 July, 2013\\
		 Stockholm, Sweden}

\begin{document}

\section{Introduction}
\label{sec:intro}
P. Gorham {\it et al.} proposed in 2008 an alternative to well established methods for the detection of ultra-high energy cosmic rays (energies greater than $10^{17}~$eV)~\cite{Gorham}. Indeed, using electron beams at ANL and SLAC facilities, they found an evidence of coherent microwave emission emitted from air shower particles. This emission, unpolarised and isotropic, was interpreted as molecular bremsstrahlung radiation (MBR), {\it i.e.}\ an emission produced by low-energy electrons scattering into the electromagnetic field of the neutral atmospheric molecules. Motivated by this experimental observation, several prototypes have been installed at the Pierre Auger Observatory to test the feasibility of this new detection technique~\cite{RGaior,MyEPS}. If such a technique were confirmed during these R\&D activities, it would improve significantly the detection efficiency of extensive air showers produced by ultra-high energy cosmic rays since its duty cycle would be close to 100\%, with a very low atmospheric attenuation. Thus, it would be possible to develop radio telescope recording the longitudinal profile of the air shower and providing a calorimetric measurement of the shower energy. Recent results from~\cite{MIDAS} disfavour the hypothesis of a quadratic scaling between the emitted power in the microwave band and the air shower energy. Also, preliminary results performed by the MAYBE collaboration~\cite{MAYBE} indicate a linear scaling with the beam energy and a lower MBR yield compared to P. Gorham {\it et al.}'s measurements. The purpose of this paper is to present a new laboratory experiment to measure precisely radio emissions produced by an electromagnetic air shower in the wide band from 2 to 20~GHz.

\section{Experimental setup}
\label{sec:setup}
The measurements are performed at the Beam Test Facility (BTF) of the LNF laboratory in Frascati (Italy), part of the DA$\phi$NE accelerator complex. It can provide electron bunches with a charge up to $10^{10}$~e$^-$/pulse, with an energy between 25 and 750~MeV and an associated bunch length from 1 to 10~ns. During the first measurement campaigns, we operated with bunch lengths of 3~ns or 10~ns, with an energy around 510~MeV. To increase the energy deposit by the air shower, the beam collides into an alumina (95\% pure Al$_{\rm 2}$O$_{\rm 3}$) target of variable thickness -- up to 6 radiations lengths X$_{\rm 0}$ -- placed just before the anechoic chamber. The inner surface of the copper anechoic chamber is covered with RF absorbers having an absorption range from 35~dB at 1~GHz to 45~dB at 6~GHz, and 50~dB for higher frequencies. The size of the chamber (2~m width $\times$ 2~m height $\times$ 4~m length) permits to be crossed by the beam exactly in its central axis. Inside the chamber, five different positions for radio receivers are possible, all of them being able to have the polarisation plane of the antenna orthogonal (cross-polarised) or parallel (co-polarised) to the beam axis. The receivers used are two log-periodic antennas -- Rohde \& Schwarz HL050 -- and two horns -- DRH20. Horn antenna calibration was done with the SATIMO StarLab calibration system~\cite{SATIMO}. All of them have a quite large frequency band, $1.7-20~$GHz and $0.85-26~$GHz, respectively. The gain of the receivers is about 12~dBi for log-periodic antennas and from 6 to 16~dBi in the case of horns. After having been amplified by about 26~dB by a wide band amplifier, the output signal is sent to an oscilloscope.

\section{Simulations and strategy for measurements}
\label{sec:simus}
A full dedicated Geant4 simulation of the setup has been performed. It aims to reproduce the propagation and the interaction of the electrons into the alumina target, and inside the anechoic chamber. It is especially useful to estimate correctly the Cherenkov radiation produced in the frequency range [1--20]~GHz and detected by the antennas in both polarisations. Indeed, inside the chamber, the Cherenkov emission will be the main contribution to the recorded signal in the microwave band. However, the Cherenkov radiation has a well defined polarisation, lying in the plane defined by the Poynting vector and the particle trajectory. Therefore, orienting the receiver polarisation plane orthogonal to the beam axis can drastically reduce the Cherenkov signal. Since the MBR radiation is expected unpolarised, the cross-polarised configuration should increase the ratio MBR signal over Cherenkov signal. Another possibility is to increase the number of radiation lengths with the alumina target since it is well known that, below energies around 20~MeV, electrons in air cannot emit Cherenkov radiation. Our preliminary data confirm that the cross-polarised signal is significantly lower than the co-polarised one. A detailed analysis of the full data set is now on going.

\section{Conclusion and outlook}
\label{sec:conclusion}
The experimental setup, the analysis strategy and the first measurements have been presented. A large data set with different configurations has been recorded during three measurement campaigns. A detailed data analysis is now the next step for this project.

\acknowledgments
The work is supported by INFN, Italy and the EC FP7 Research Infrastructure projects (HadronPhysics3, GA n. 283286 and AIDA, GA 262025). The work is also supported by the MSMT CR grant LE13012.

\end{document}